\documentclass[aps,twocolumn,nofootinbib,showpacs,amsmath,superscriptaddress,prl,reprint,preprintnumbers]{revtex4-2}

\usepackage{graphicx}
\usepackage{epstopdf}
\usepackage{bm}
\usepackage{epsfig}
\usepackage{graphics}
\usepackage{xspace}
\usepackage{tikz}
\usepackage{xcolor}
\usepackage[normalem]{ulem}
\usetikzlibrary{calc}
\usepackage{orcidlink}

\usepackage{subfigure}

\newcommand{\ord}[1]{\mathcal{O}(#1)}

\def\OMIT#1{}

\definecolor{mathematicaBlue}{rgb}{0.368417, 0.506779, 0.709798}
\definecolor{mathematicaOrange}{rgb}{0.880722, 0.611041, 0.142051}
\definecolor{mathematicaGreen}{rgb}{0.560181, 0.691569, 0.194885}
\definecolor{mathematicaRed}{rgb}{0.922526, 0.385626, 0.209179}
\definecolor{darkred}{rgb}{0.7, 0,0}
\definecolor{darkblue}{rgb}{0,0,0.7}
\definecolor{darkgreen}{rgb}{0, 0.6,0}

\newcommand{\msbar}{\overline{\textrm{MS}}}

\newcommand{\nn}{\nonumber}
\newcommand{\df}{{\rm d}}

\newcommand{\bea}{\begin{eqnarray}}
\newcommand{\eea}{\end{eqnarray}}

\newcommand{\gsim}{\mathrel{\rlap{\lower4pt\hbox{\hskip1pt$\sim$}}\raise1pt\hbox{$>$}}}

\newcommand{\Om}{\Omega_1^\rho}
\newcommand{\Th}{\Theta_1}

\newcommand{\apeak}{a_{\text{peak}}}

\usepackage{makecell}
\setcellgapes{2pt}
\makegapedcells

\newcommand{\be}{\begin{equation}}
\newcommand{\ee}{\end{equation}}

\begin{document}

\title{\boldmath A Precise Determination of $\alpha_s$ from the Heavy Jet Mass Distribution}

\author{Miguel A.\ Benitez\orcidlink{0000-0002-6939-2677}}
\affiliation{Departamento de F\'isica Fundamental e IUFFyM, Universidad de Salamanca, E-37008 Salamanca, Spain}

\author{Arindam Bhattacharya\orcidlink{0000-0002-4457-8926}}
\affiliation{Department of Physics, Harvard University, Cambridge, MA 02138, USA}

\author{Andr\'e H.\ Hoang\orcidlink{0000-0002-8424-9334}}
\affiliation{University of Vienna, Faculty of Physics, Boltzmanngasse 5, A-1090 Wien, Austria}

\author{\\Vicent Mateu\orcidlink{0000-0003-0902-5012}}
\affiliation{Departamento de F\'isica Fundamental e IUFFyM, Universidad de Salamanca, E-37008 Salamanca, Spain}

\author{Matthew D.\ Schwartz\orcidlink{0000-0001-6344-693X}}
\affiliation{Department of Physics, Harvard University, Cambridge, MA 02138, USA}

\author{Iain W.\ Stewart\orcidlink{0000-0003-0248-0979}}
\affiliation{University of Vienna, Faculty of Physics, Boltzmanngasse 5, A-1090 Wien, Austria}
\affiliation{Center for Theoretical Physics -- a Leinweber Institute, Massachusetts Institute of Technology, Cambridge, MA 02139, USA}

\author{Xiaoyuan Zhang\orcidlink{0000-0002-2090-2381}\vspace{0.2cm}}
\affiliation{Department of Physics, Harvard University, Cambridge, MA 02138, USA}
\affiliation{Center for Theoretical Physics -- a Leinweber Institute, Massachusetts Institute of Technology, Cambridge, MA 02139, USA}

\begin{abstract}

A global fit for $\alpha_s(m_Z)$ is performed on available $e^+e^-$ data for the heavy jet mass distribution. The state-of-the-art theory prediction includes $\mathcal{O}(\alpha_s^3)$ fixed-order results, N$^3$LL$^\prime$ dijet resummation, N$^2$LL Sudakov shoulder resummation, and a first-principles treatment of power corrections in the dijet region. Theoretical correlations are incorporated through a flat random-scan covariance matrix. The global fit results in $0.1148^{+ 0.0015}_{-0.0022}$, compatible with similar determinations from thrust and \mbox{$C$-parameter}.
Dijet resummation is essential for a robust fit, as it engenders insensitivity to the fit-range lower cutoff; without resummation the fit-range sensitivity is overwhelming.
In addition, we find evidence for a negative power correction in the trijet region if and only if Sudakov shoulder resummation is included.

\end{abstract}

\pacs{12.38.Bx, 12.38.Cy, 12.39.St, 24.85.+p}

\preprint{MIT-CTP 5840, UWTHPH 2025-7}

\maketitle
One of the main targets of precision QCD is the measurement of the strong coupling at colliders. In this endeavor, studies of $e^+e^-$ event shapes such as thrust and heavy jet mass (HJM) have led to enormous progress in our understanding of the interface between perturbative and non-perturbative QCD. Intriguingly, attempts to extract $\alpha_s$ from HJM data consistently produced values well below that from thrust \cite{Salam:2001bd,MovillaFernandez:2001ed,Gardi:2002bg,Dinsdale:2004zw,Becher:2008cf,Dissertori:2009ik,Chien:2010kc,Alioli:2012fc,Banfi:2023mes}, and below the PDG
average~\cite{ParticleDataGroup:2024cfk}. In recent years, a number of differences between HJM and thrust were observed. First, HJM (in contrast to thrust) has a left-Sudakov shoulder due to large logarithms as $\rho\to 1/3$ from below~\cite{Bhattacharya:2022dtm,Bhattacharya:2023qet}. Second, in a recent study of soft emissions from three-parton configurations~\cite{Luisoni:2020efy,Caola:2021kzt, Caola:2022vea, Nason:2023asn,Nason:2025qbx} it was suggested that these yield a positive non-perturbative power correction for thrust, shifting the distribution to the right, but a negative one for HJM, shifting the distribution leftward. As we will discuss, there is evidence that the Shoulder and the negative shift are intertwined. Third, power corrections in moments of the HJM distribution depend on different non-perturbative parameters than the tail OPE~\cite{Hoang:2025uaa}, while in thrust the parameters are the same.
In this study, we incorporate all these effects into an improved fit. In addition, we include a more sophisticated treatment of correlated theoretical uncertainties using a random-scan covariance matrix which is added to the experimental one in the fits. This treatment of uncertainties
leads to results robust across a wide range of fit regions with $\chi^2/{\rm dof} \approx 1$.

\begin{figure}[t!]
\centering
\begin{tikzpicture}
\node[anchor=south west, inner sep=0] (img) at (0,0) {\includegraphics[width=\columnwidth]{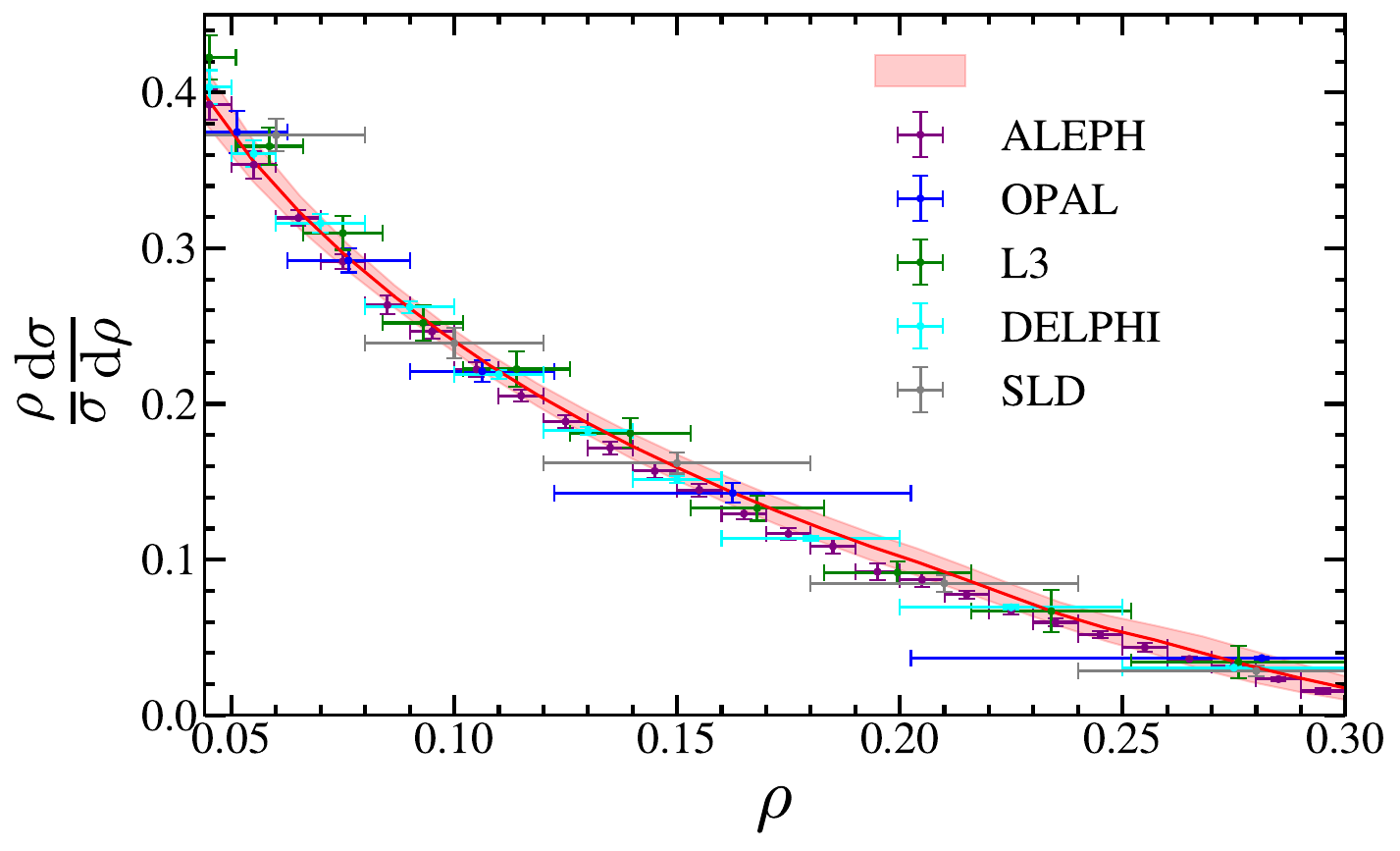}};
\node[black, anchor=west,scale=1] at (6.1, 4.8) {theory};
\end{tikzpicture}
\caption{Comparison of the theory prediction to
data from five experiments at $Q=91.2$\,GeV. Horizontal error bars indicate the sizes of the experimental bins.}
\label{fig:vslep}
\end{figure}

{\bf Experimental data:} Data on the heavy jet mass cross section are given as binned distributions by the {\sc aleph}~\cite{Heister:2003aj}, {\sc delphi}~\cite{Abreu:1996na,Abreu:1999rc,Abdallah:2003xz}, {\sc jade}~\cite{MovillaFernandez:1997fr}, {\sc l3}~\cite{Adeva:1992gv,Achard:2004sv} {\sc opal}~\cite{Ackerstaff:1997kk,Abbiendi:1999sx,Abbiendi:2004qz}, and {\sc sld}~\cite{Abe:1994mf} collaborations. These measurements comprise 50 datasets at $35$ different center-of-mass energies between $35$\,GeV and $207$\,GeV for a total of 700 experimental data points. Data at $Q=91.2$\,GeV is shown compared to our best theory result in Fig.~\ref{fig:vslep}.

Experimental measurements are generally provided with statistical and systematic experimental uncertainties. To treat correlations among the latter, we employ the LEP QCD working group~\cite{Heister:2003aj,Abbiendi:2004qz} minimal overlap model. We denote the statistical and systematical uncertainty for each data point (bin) by $\Delta_i^{\rm stat}$ and $\Delta_i^{\rm sys}$, respectively. We denote by $D_i$ to which of the 50 datasets, defined by energy and experiment, data-point $i$ belongs. Then, the minimal-overlap experimental covariance matrix is
\begin{equation}
\sigma_{ij}^{\rm exp} =
\delta_{ij} (\Delta_i^{\rm stat})^2 +
\delta_{D_iD_j} {\rm min}(\Delta_i^{\rm sys},\Delta_j^{\rm sys})^2\,.
\label{sigmaexp}
\end{equation}
That is, $ \sigma_{ij}^{\rm exp}$ is block-diagonal by dataset. This model assumes a positive correlation of systematic uncertainties within each dataset.

Each dataset for HJM has a peak at some value of $\rho$. These values of the peak position, which are described by \mbox{$\rho_\text{peak} = 0.003 + \apeak/Q$} with $\apeak = 1.56$\,GeV (cf.~supplement Fig.~\ref{fig:peaks}), provide a quantitative indication of the scale at which
the distribution becomes non-perturbative. We thus restrict our fits to $\rho\, Q \ge 3 \apeak$ to keep safely away from the fully non-perturbative region. We also impose $\rho \le 0.3$ on our fits to avoid analogous non-perturbative effects in the shoulder region as well as unknown higher fixed-order contributions which are relatively more important for higher values of $\rho$.
There are 451 experimental data points whose entire bin satisfies $3\apeak/Q \le \rho \le 0.3$.

\vspace*{2mm}
{\bf HJM Distribution:}
The HJM differential cross section factorizes in the dijet limit into a global hard factor $H_{\text{dij}}$ times the two-dimensional convolution of two \mbox{one-dimensional} jet functions $J_{1,2}$, and the two-dimensional soft and shape functions~\cite{Catani:1991bd,Korchemsky:1999kt,Fleming:2007qr,Fleming:2007xt,Schwartz:2007ib,Hoang:2025uaa}:
\begin{equation}\label{eq:Xsec}
\df \sigma_{\text{dij}} =
H_{\text{dij}}\times J_1 \times J_2\otimes S_{1,2}
\otimes F^{\Xi}_{1,2}(\Om)\,.
\end{equation}
Here,
the hard, jet and soft functions include large-log resummation, and
the shape function $F^{\Xi}_{1,2}$ is given by matrix elements of QCD operators. For our fits it suffices to take the shape function to depend on a single non-perturbative dijet parameter $\Om$, which is of ${\cal O}(\Lambda_{\rm QCD})$, and defined in the R-gap scheme that cancels the leading renormalon of the partonic soft function, with reference scale $2\,{\rm GeV}$~\cite{Abbate:2010xh,Benitez:2024nav}.

Around the symmetric trijet limit, the distribution also factorizes as~\cite{Bhattacharya:2022dtm,Bhattacharya:2023qet}
\begin{equation}\label{eq:Xsecsh}
\df \sigma_{\text{sh}}^{\text{pert}} = H_{\text{sh}}\times J_1\times J_2 \times J_3 \otimes S_{1,2,3}\,.
\end{equation}
While dijet resummation can be done analytically, shoulder resummation requires a numerical Fourier transform from position to momentum space
which is computationally expensive. Although no first-principles study has been performed of power corrections around the symmetric trijet configuration, we model those corrections with a non-perturbative shift parameter. That is, we take
\begin{equation}
\frac{\df\sigma_{\text{sh}}}{\df \rho}(\rho)
=\frac{\df\sigma^\text{pert}_{\text{sh}}}{\df \rho}\biggl(\rho -\frac{\Th}{Q}\biggr)\,.
\end{equation}
Here $\Th <0$ corresponds to a negative shift in the trijet region. This is natural, since the shoulder expansion is around $r=1/3-\rho$, so a positive/physical non-perturbative shift in $r$ pushes the distribution to smaller $\rho$, and thus has $\Th <0$.

Matching between the dijet, fixed-order and shoulder regions is done by writing the full cross section as
\begin{equation}
\df\sigma = \Bigl[\df\sigma_{\text{dij}} - \df\sigma_{\text{dij}}^{\text{sing}}\Bigr] + \df\sigma_{\text{FO}}
+ \Bigl[\df\sigma_{\text{sh}} - \df\sigma_{\text{sh}}^{\text{sing}}\Bigr]\,.
\end{equation}
Here $\df\sigma_{\rm FO}$ is fixed-order, known currently to ${\cal O}(\alpha_s^3)$~\cite{Catani:1996jh,Catani:1996vz,Weinzierl:2008iv,Ridder:2014wza,DelDuca:2016csb,DelDuca:2016ily,Aveleira:2025svg}, while $\df\sigma_{\rm dij}$ and $\df\sigma_{\rm sh}$ resum large logarithms. $\df\sigma^{\text{sing}}_X$ refers to $\df\sigma_X$
with the jet, soft and hard scales set equal.
With equal scales, each distribution reduces to a set of fixed-order terms involving $\ln^k(\rho)$ or $\ln^k(1/3-\rho)$. The differences in brackets are power suppressed in regions where their logarithms are not large. When logarithms are large, the $\df\sigma^{\text{sing}}_X$
subtract terms from $\df\sigma_{\rm FO}$ and avoid double counting. In practice, all regions overlap, so the interpolation between regions is handled by modulating the hard, jet and soft scales with profile functions $\mu_i (\rho)$.
Profile functions for HJM have been discussed extensively elsewhere~\cite{Hoang:2014wka,Hoang:2025uaa,Bhattacharya:2023qet}. We summarize our choice of profiles in the supplemental material, providing the variation intervals of their parameters in Table~\ref{tab:profParams}.

To compare to the experimental data, the theory prediction should be integrated over each experimental bin. A practical implementation of this integration is by using the cumulative cross section in which the renormalization scales are set to the profiles evaluated at the center of the bin~\cite{Abbate:2010xh}. This avoids an expensive numerical integration over the dijet region, cancels large contributions from the peak region, and is as accurate as a direct integration within the order in perturbation theory one is working.

\vspace*{2mm}
{\bf Fit Procedure:}
There is no standard way to estimate theoretical uncertainties, or to include theoretical uncertainties in fits to experimental data. A typical procedure is to perform a fit by minimizing $\chi^2$ using the experimental covariance matrix for various choices of theory parameters and estimating the theory uncertainty from the variation of the fit results~\cite{Dissertori:2007xa,Heister:2003aj}. Such a procedure can bias the results. For instance, if a single bin has nearly zero experimental uncertainty but enormous theory uncertainty, it will dominate the fits even if the agreement between theory and experiment is not good. Therefore, in general, it is critical to incorporate theory uncertainties into the fitting procedure to get sensible results.

Since theory uncertainties are assessed through renormalization scale variation, they are not Gaussian and highly correlated. To account for the correlations we employ a {\it flat random scan}. In such a scan, a range is first defined for each of the $k$ theory uncertainty parameters. Next, $M= 5000$ sets of the $k\le 16$ parameters are generated by independently sampling each parameter from a uniform distribution within its specified range, ensuring a uniform (flat) measure across the $k$-dimensional parameter space. Among these $M$ sets, we include $2k$ non-random sets, given by varying one parameter to its maximal or minimal value with the other parameters fixed at their default values. For each of the $M$ sets, the theory prediction is computed for each data-point $x_i$. We define $\bar x_i= (x_i^{\text{max}} + x_i^{\text{min}})/2$ and $\Delta^{\rm theo}_i = (x_i^{\text{max}} - x_i^{\text{min}})/2$ where $x_i^{\text{max}}$ and $x_i^{\text{min}}$ are the maximum and minimum among the $M$ predictions for each $x_i$. Then $\bar x_i$ can be interpreted as the theory prediction for bin $i$ and $\Delta^{\rm theo}_i$
as its $1$-$\sigma$ uncertainty. Note that the theory predictions ${\bar x}_i$ do not correspond to a particular choice of the theory parameters, but are not far from the default set.

\begin{figure}
\centering
\includegraphics[width=\columnwidth]{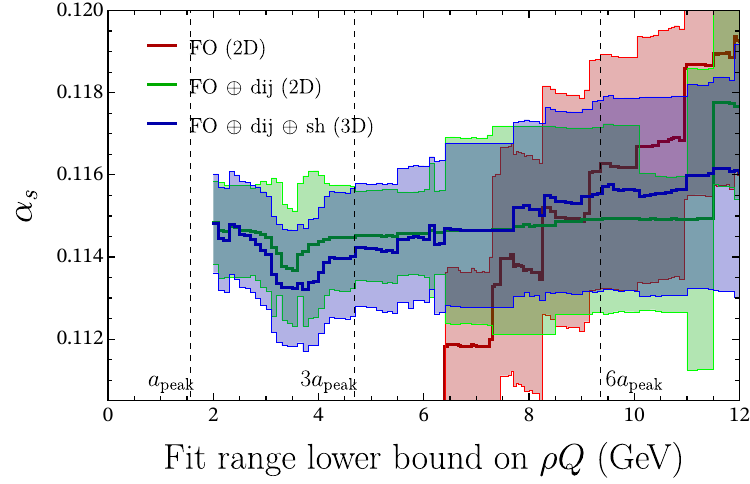}
\caption{Results for $\alpha_s$ using a fit range $a/Q \le \rho \le 0.3$ as a function of $a$. Two 2D fits ($\alpha_s,\Om$) using fixed-order (FO) or FO with dijet resummation (dij) are shown, as well as a 3D fit ($\alpha_s,\Om,\Th$) that includes Sudakov shoulder (sh) resummation as well.
The dashed lines represent integer multiples of the non-perturbative boundary $Q\rho = \apeak$.}
\label{fig:lowercut}
\end{figure}

The
correlation coefficient $r_{ij}$ among bins $i$ and $j$ is then computed with the standard formula
\begin{align}\label{eq:theo-cor}
r_{ij}^{\text{theo}} = \frac{\langle(x_i - \bar x_i) (x_j - \bar x_j)\rangle}
{\sqrt{\langle(x_i - \bar x_i)^2\rangle}\sqrt{\langle(x_j - \bar x_j)^2\rangle}}\,,
\end{align}
where $\left< y \right>$ represents the average of the quantity $y$ over the $M$ theory parameter sets.
The theory covariance matrix results from scaling the correlation coefficient matrix by the 1-$\sigma$ uncertainties:
\begin{align}\label{eq:theo-cov}
\sigma_{ij}^{\rm theo} = \Delta_i^{\rm theo}\,\Delta_j^{\rm theo}\,r_{ij}^{\rm theo}\,.
\end{align}
The total covariance matrix is the sum of the theoretical and experimental ones:
$\sigma_{ij}^{\rm tot} = \sigma_{ij}^{\rm theo} + \sigma_{ij}^{\rm exp}$, with $\sigma^{\rm exp}_{ij}$ given in Eq.~\eqref{sigmaexp}.
The $\chi^2$ is then:
\begin{align}\label{eq:chi2}
\chi^2 = \sum_{i,j=1}^{N_{\rm bins}} (\bar x_i - x_i^{\rm exp})\,
(\bar x_j - x_j^{\rm exp})\,(\sigma_{\rm tot}^{-1})_{ij}\,.
\end{align}
This $\chi^2$ is computed for each value of $(\alpha_s,\Om)$ or $(\alpha_s,\Om,\Th)$ in a regular grid. Then, the minimum is found from a 2D or 3D interpolation of the $\chi^2$ values. Note that we compute the 3D grid with both positive and negative values of $\Th$, and let the $\chi^2$ determine its sign. The 1-$\sigma$ uncertainty in $\alpha_s$ (for a fixed fit range) is computed by profiling, i.e.\ minimizing $\chi^2$ over the non-perturbative parameters for each value of $\alpha_s$ and extracting the range for which $\Delta \chi^2=1$. Note that the $\chi^2$ function includes both theoretical and experimental uncertainties, which can no longer be disentangled.

\begin{figure}[t!]
\centering
\resizebox{0.9\columnwidth}{!}{\begin{tikzpicture}
\node[anchor=south west, inner sep=0] (img) at (0,0) {\includegraphics[width=\columnwidth]{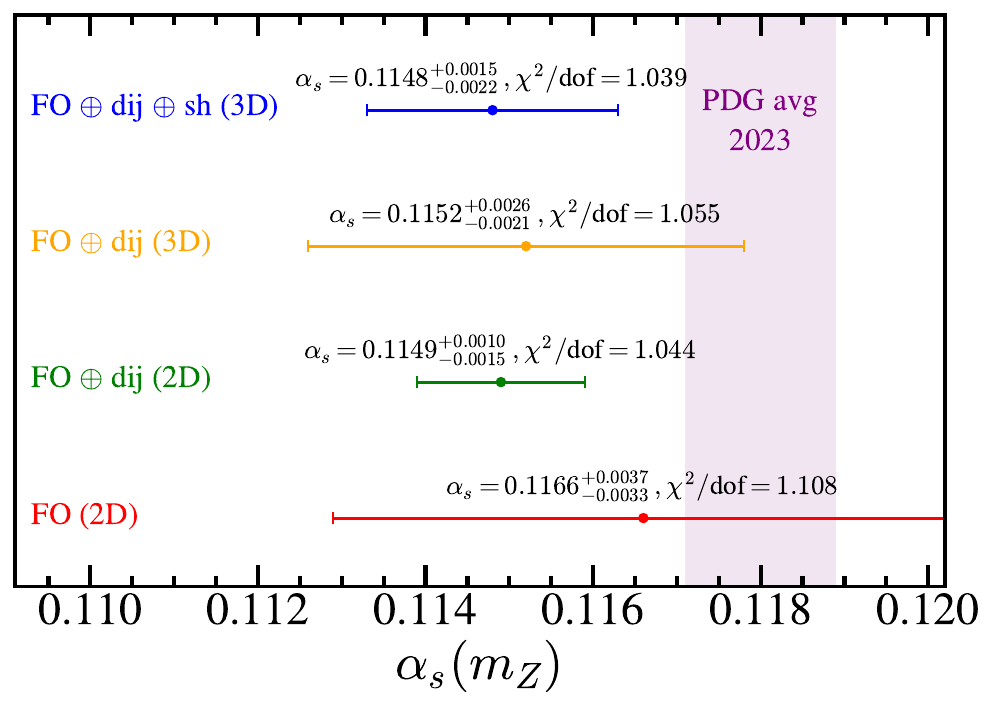}};
\end{tikzpicture}}
\caption{Results for $\alpha_s$ with various levels
of theoretical sophistication.}
\label{fig:alphasfits}
\end{figure}

We show in Fig.~\ref{fig:lowercut} results of the fits for $\alpha_s$ using three theoretical predictions. The fit range is over \mbox{$a/Q \le \rho \le 0.3$} for various choices of $a$. As $a$ increases, experimental data-points become excluded when the entire bin is not within the fit range. The first two fits are two-parameter fits to $\alpha_s$ and a single non-perturbative shape parameter $\Om$ using 1) NNLO perturbation theory (FO), and 2) FO matched to dijet resummation only. Resummed results use profile parameters to transition smoothly between the dijet and fixed-order regions. We see that the fixed-order result is very sensitive to the fit range, making it impossible to extract a sensible value of $\alpha_s$ without an arbitrary choice of fit range. With resummation included, the fit value is remarkably insensitive to the fit range. Indeed, nearly the same value results from $\rho Q > 2.5\,$GeV with 525 data points and $\rho Q > 11\,$GeV with 306 data points. For the third fit 3) Sudakov shoulder resummation is also incorporated and an additional non-perturbative parameter $\Th$ is included to parametrize power corrections in the trijet region.

\begin{figure}[t!]
\centering
\includegraphics[width=\columnwidth]{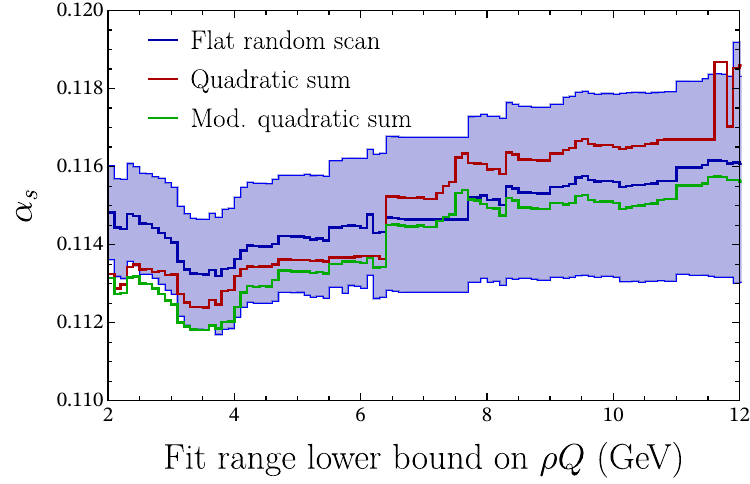}
\caption{Results for $\alpha_s(m_Z)$, in the best-fit setup, as a function of the lower bound of the fit range, for three different methods of determining the theory covariance matrix. }
\label{fig:comparisonFits}
\end{figure}

\vspace*{2mm}
{\bf Alternate theory covariance:} In order to assess the impact of the treatment of theoretical uncertainties and the corresponding correlations, we have used a prescription different from the flat random scan and repeated the fit for $\alpha_s(m_Z)$. In particular, we consider the variation of each profile parameter independently, including the correlations it induces for uncertainties in different bins of $\rho$. The full covariance matrix for perturbative uncertainties is then obtained from the quadratic sum of these individual uncertainties. Furthermore, we also consider a modified quadratic sum method for reference, where we take the quadratic sum of individual profile variations first, and then form the correlation matrix. In Fig.~\ref{fig:comparisonFits} we show a comparison of the three methods in the $\text{FO}\oplus \text{dij}\oplus \text{sh 3D}$ fit, which yield compatible results for $\alpha_s(m_Z)$. The overall size of the $\alpha_s$ uncertainties in the quadratic sum method is similar to the flat random scan, however there are smaller uncertainties for some bins, leading to a preference for the flat random scan method. We use the difference between the quadratic sum and flat random scan correlation methods as an additional systematic uncertainty.

\vspace*{2mm}
{\bf Results:}
To extract a value of $\alpha_s$ from these fits, we perform a weighted average over the choices for the lower bound of $a \le \rho Q$ with a weight $w_a = \sigma_a^{-2}(\sum_a \sigma_a^{-2})^{-1}$ with $\sigma_a$ the uncertainty for a given fit range. For the resummed results we consider $a \in [3 \apeak, 6 \apeak]$ in steps of $0.5 \apeak$. The results are shown in Fig.~\ref{fig:alphasfits} (a full breakdown of uncertainties is given in the supplemental material, Table~\ref{tab:fits}). Using the full theoretical model, including dijet, fixed-order and shoulder, and three parameters $\alpha_s$, $\Om$ and $\Th$ gives, after marginalizing over $\Om$ and $\Th$
\begin{align} \label{fit}
\alpha_s(m_Z) &= 0.1148\,
{}^{+ 0.0005}_{-0.0010}\,\text{(th+exp)}
~{}^{+ 0.0011}_{-0.0017}\,(\Om)
\nn \\
&\hspace{-10pt}~ {}^{+ 0.0002}_{-0.0001}\,(\Th)
~{}^{+0.0005}_{-0.0005}\,\text{(fit range)}~{}^{+0.0008}_{-0.0008}\,\text{(fit method)} \nonumber\\
&= 0.1148^{+ 0.0015}_{-0.0022}\,,
\end{align}
with a $\chi^2/{\rm dof} = 1.039$. The results for $\Om$ and $\Th$ are
\begin{equation} \label{eq:NPfitresults}
\Om = 0.61 \pm 0.08\, \text{GeV},\quad \Th =-0.46 \pm 0.17\, \text{GeV}\,,
\end{equation}
where the uncertainties correspond to $\Delta \chi^2=1$. Here, the ``th+exp" includes experimental systematic and statistical uncertainty as well as perturbative uncertainty from the variation of the 17 theory parameters, all encoded in the total covariance matrix. The dominant uncertainty in $\alpha_s(m_Z)$ is due to non-perturbative effects in the dijet region encoded in $\Om$. (The breakdown of uncertainties for individual fit ranges can be found in the left panel of Fig.~\ref{fig:errorBudgetRandomScan}.) In Eq.~\eqref{fit}, the fit-range uncertainty, given by the standard deviation among the best-fit values for the various lower bounds considered, is subdominant. Using the results from Fig.~\ref{fig:comparisonFits}, we find $\alpha_s=0.1140\pm 0.0014$, $\Omega_1=0.61\pm 0.04$ GeV and $\Theta_1=-0.42\pm 0.12$ GeV
for the quadratic sum after averaging over the fit range. We include the difference between the central values from the flat random scan and quadratic sum methods as an additional uncertainty. Fitting $\alpha_s(m_Z)$ and $\Om$ using only $\df\sigma = \df\sigma_{\rm FO}$, and taking slightly higher lower cutoffs, $a \in [5 \apeak, 8 \apeak]$, gives $\alpha_s= 0.1166^{+0.0037}_{-0.0033}$. In this case, the fit range uncertainty increases significantly, as can be seen clearly in Fig.~\ref{fig:lowercut}.

\begin{figure}[t!]
\centering
\includegraphics[width=\columnwidth]{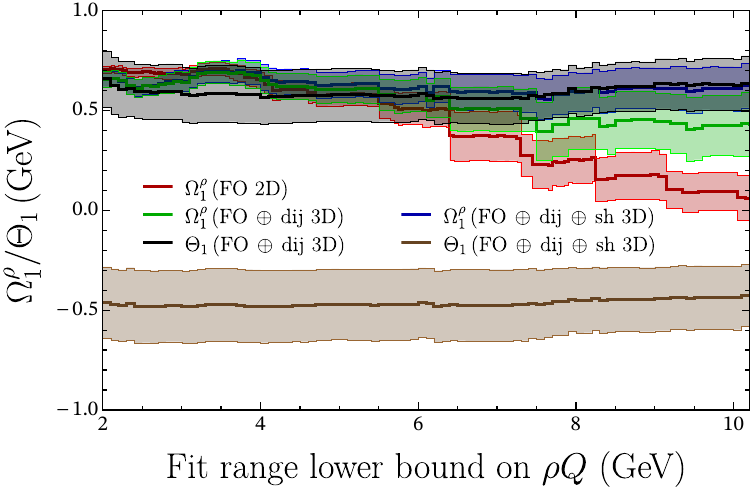}
\caption{Results for the non-perturbative parameters with various levels
of theoretical sophistication. 2D fits are to $\alpha_s$ and $\Om$ while 3D fits also fit to $\Th$. The $\Om$ band in a 2D fit to FO + dij is similar to the blue band and not shown.}
\label{fig:omega}
\end{figure}

It is instructive to look also at the values of the non-perturbative parameters in Eq.~(\ref{eq:NPfitresults}), which are shown in Fig.~\ref{fig:omega}. Our fits indicate that when using fixed order or dijet resummation, the data prefers a positive power correction
which shifts the distribution rightwards, to larger~$\rho$. We also find that when shoulder resummation is included, the data supports a rightward shift for small $\rho$ (positive $\Om$) transitioning to a leftward shift in the far-tail (negative $\Th$). In contrast, without shoulder resummation the 3D fit gives positive $\Th$ and $\Om$. This contrasts with recent work~\cite{Nason:2023asn,Nason:2025qbx,Caola:2021kzt,Caola:2022vea} which suggested that HJM should have a leftward shift throughout.\footnote{The computation in~\cite{Nason:2023asn,Nason:2025qbx} is based on a perturbative calculation using a three-parton dipole model of non-perturbative shift function $\zeta(\rho)$. Their fits did not include resummation. No fits to $\rho$ alone were given in~\cite{Nason:2025qbx}, only fits combined with other event shapes.} The emergence of a region with a negative power correction validates the importance of including shoulder resummation and motivates further study of power corrections in HJM.

The exact treatment of hadron masses in the experimental measurements affects the value of the dijet parameter $\Omega_1$ for HJM and other event shapes~\cite{Salam:2001bd,Mateu:2012nk}. We assume that the data are all based on a common procedure for dealing with hadron masses, hence in our theory description non-perturbative corrections can be parametrized by the same $\Omega_1^\rho$. Hadron mass effects are, however, relevant when comparing our value $\Omega_1^\rho=0.61 \pm 0.08\, \text{GeV}$ to measurements from other event shapes, such as $\Omega_1^\tau/2 = \Omega_1^R =0.31\pm 0.05\,{\rm GeV}$~\cite{Benitez:2024nav} from thrust in the R-gap subtraction scheme. These are expected to differ since thrust and HJM belong to different universality classes~\cite{Mateu:2012nk}. Neglecting hadron mass effects one would expect $\Omega_1^\rho = \Omega_1^\tau/2$, but including them gives an expectation from Monte Carlo simulations that $\Omega_1^\rho \simeq \Omega_1^\tau$. This is nicely compatible with our result.

\vspace*{0.2cm}
{\bf Cross checks}
Next we carry out some tests of the robustness of our analysis. We now consider only the full theory prediction (FO\,$\oplus$\,dij\,$\oplus$\,sh\,$\oplus\,\Om\oplus\Th$) on subsets of the 410 datapoints with $6\,{\rm GeV}/Q \le \rho \le 0.3$ for which $\alpha_s(m_Z) = 0.1144 \pm 0.0018$ and $\chi^2=427$.
For results quoted from these analyses we do not include fit range or fit method uncertainty.

To assess sensitivity to missing effects from the mass of the $b$ quark, we consider excluding the lower energy data $Q$, following~\cite{Abbate:2010xh,Abbate:2012jh}. Using the subset with $Q \ge 90$\,GeV (355 points) gives $\alpha_s = 0.1145 \pm 0.0020$. Taking $Q \le 100$\,GeV (99 points) gives $\alpha_s = 0.1163 \pm 0.0021$. To address concerns about the quality of the {\sc delphi} data~\cite{Wicke:1999zz,Nason:2025qbx}, we refit removing all {\sc delphi} datasets, giving $\alpha_s = 0.1142 \pm 0.0021$, and also fit using~\cite{Wicke:1999zz}
which gives $\alpha_s=0.1143 \pm 0.0020$ (and increases $\chi^2$ by $14.1$). These results are consistent with our global fits.

Finally we carry out analyses to emphasize the importance of handling correlations properly. First, we note that if we drop the off-diagonal terms in $\sigma_{ij}^{\rm exp}$, we find
$\alpha_s = 0.1137 \pm 0.0018$, within our uncertainties. However, if we drop the important off-diagonal terms in $\sigma_{ij}^{\rm theo}$ we get
$\alpha_s = 0.1125 \pm 0.0026$, which is a bigger effect. This suggests that inclusion of the full theory correlations is important to obtain proper fit results.

\vspace*{0.2cm}
{\bf Conclusions}
We have provided a comprehensive analysis of available data on heavy jet mass. Innovations include improved treatments of the dijet/OPE and trijet/shoulder region, inclusion of theory correlations during the fitting, and careful attention to the range of data used for fitting. We found that, with resummation, fits are minimally sensitive to the fit range, in contrast to fixed-order perturbation theory which leads to an essentially linear dependence of the extracted $\alpha_s$ on the lower bound of the fit range. We find evidence for a negative power correction in the tail of the distribution only if Sudakov shoulder resummation is included.
Our extracted value is $0.1148^{+ 0.0015}_{-0.0022}$,
which is compatible with the result determined from thrust~\cite{Benitez:2024nav,Abbate:2010xh} and \mbox{$C$-parameter}~\cite{Hoang:2015hka},
and compatible to the world average~\cite{ParticleDataGroup:2024cfk} at the 1.8-$\sigma$ level.

\vspace*{0.2cm}
{\bf Acknowledgments:}
We would like to thank F.~Caola, P.~Monni, P.~Nason, F.~Tackmann and G.~Zanderighi for helpful discussions. This work was partially supported in part by FWF Austrian Science Fund under the Project No.~P32383-N27, the U.S.\ Department of
Energy Office under the Contracts DE-SC0011090 and DE-SC0013607, the Spanish MICIU/AEI/10.13039/501100011033 grant No.\ PID2022-141910NB-I00, the JCyL grant SA091P24 under program EDU/841/2024 and the Research in Teams Program under the project ``Improving the Precision of Predictions for Event-Shape Distributions" of the Erwin Schr\"odinger International Institute for Mathematics and Physics. MB was supported by a JCyL scholarship funded by the regional government of Castilla y Le\'on and European Social Fund, 2022 call. IS was also supported by the Simons Foundation through the Investigator grant 327942. XYZ was also supported by the MIT Pappalardo
Fellowship.
We thank the Erwin Schr\"odinger International Institute for Mathematics and Physics for hospitality during the 2023 Thematic Programme ``Quantum Field Theory at the Frontiers of the Strong Interaction''.
The computations in this paper were run on the FASRC Cannon cluster supported by the
FAS Division of Science Research Computing Group at Harvard University.

\bibliography{thrust3}

\setcounter{equation}{0}
\renewcommand{\theequation}{S-\arabic{equation}}
\clearpage

\newpage
\begin{widetext}
\section*{Supplemental Material}

{\bf Profile Functions:}
The region of the HJM distribution where dijet resummation is applied is governed by three renormalization scales: hard ($\mu_H$), jet ($\mu_J$), and soft ($\mu_S$). A detailed discussion surrounding the treatment of these scales can be found in Ref.~\cite{Hoang:2025uaa}. At this point, we only briefly review their functional form.

\vspace*{1mm}
\begin{table*}[b]
\centering
\begin{minipage}[b]{0.3\textwidth}
\centering
\begin{tabular}{|c c c|}
\hline
\textbf{Parameter} & \textbf{Default} & \textbf{Range} \\ \hline
$\mu_S(0)$ & 1.1\,GeV & -- \\
$R(0)$ & 0.7\,GeV & -- \\
$n_0$ & 2 & [1.5, 2.5] \\
$n_1$ & 10 & [8.5, 11.5] \\
$t_2$ & 0.25 & [0.225, 0.275] \\
$t_s$ & 0.40 & [0.375, 0.425] \\
$r_s$ & 2 & [1.33, 3] \\
\hline
\end{tabular}
\end{minipage}
\hfill
\begin{minipage}[b]{0.3\textwidth}
\centering
\begin{tabular}{|c c c|}
\hline
\textbf{Parameter} & \textbf{Default} & \textbf{Range} \\ \hline
$e_J$ & 0 & [$-1$, 1] \\
$v_h$ & $-1~\;\,$ & [$-2$, 0] \\
$n_s$ & 0 & [$-1$, 1] \\
$v_s$ & 0 & [$-1$, 1] \\
$v_j$ & 0.5 & [0.4, 0.6] \\
$\rho_{L1}$ & 0.20 & [0.17, 0.23] \\
$\rho_{L2}$ & 0.28 & [0.25, 0.31] \\
\hline
\end{tabular}
\end{minipage}
\hfill
\begin{minipage}[b]{0.3\textwidth}
\centering
\begin{tabular}{|c c c|}
\hline
\textbf{Parameter} & \textbf{Default} & \textbf{Range} \\ \hline
$\rho_{R1}$ & 1/3 & -- \\
$\rho_{R2}$ & 0.342 & -- \\
$s_{32}$ & 0 & [$-250$, 250] \\
$s_{34}$ & 0 & [$-7$, 7] \\
$\epsilon_2$ & 0 & [$-1$, 1] \\
$\epsilon_3$ & 0 & [$-1$, 1] \\
& & \\
\hline
\end{tabular}
\end{minipage}
\caption{Dijet and shoulder profile-function parameters, with the corresponding ranges in which they are varied during the random scan.}
\label{tab:profParams}
\end{table*}

The hard scale is defined as
\begin{equation}
\label{eq:hardScale}
\mu_H = 2^{v_h} Q\,.
\end{equation}
Compared to Ref.~\cite{Hoang:2014wka}, Eq.~(\ref{eq:hardScale}) has been adapted such that its definition is compatible with Ref.~\cite{Bhattacharya:2023qet}. The variation of the profile parameter $v_h$, together with all other profile-parameter ranges, is listed in Table~\ref{tab:profParams}.
The soft scale reads
\begin{eqnarray}
\label{eq:softScale}
\mu_S(\rho) = \! \left\{\begin{array}{lrcl}
\mu_S(0)& 0 &\le& \rho < t_0
\\
\zeta(\mu_S(0),0,0,r_s \mu_H,t_0,t_1,\rho)\qquad &t_0 &\le& \rho < t_1
\\
r_s \mu_H \rho & t_1 &\le& \rho < t_2
\\
\zeta(0,r_s \mu_H,\mu_H,0,t_2,t_s,\rho)\quad &t_2 &\le& \rho < t_s
\\
\mu_H & t_s &\le& \rho < 0.5
\end{array}
\right.\!\!.
\end{eqnarray}
In Eq.~(\ref{eq:softScale}), between $t_0$ and $t_1$ we transition from the non-perturbative
to the resummation region. In the resummation region, bounded by $t_1 < \rho < t_2$, $r_s$ corresponds to the linear slope of the soft scale. The transition from the resummation toward the fixed-order region occurs between $t_2$ and $t_s$. The latter defines the value of $\rho$ at which all renormalization scales become equal. The definition of the function $\zeta(a_1,b_1,a_2,b_2,t_1,t_2,t)$ can be found in Ref.~\cite{Hoang:2014wka}.
The jet scale is defined as
\begin{eqnarray}
\label{eq:jetScale}
\! \!\mu_J(\rho) =\! \left\{\!
\begin{array}{ll}
\left[ 1+e_J (\rho-t_s)^2 \right] \sqrt{\mu_H \mu_S(\rho)} \qquad & \rho \le t_s
\\
\mu_H & \rho > t_s
\end{array}
\right.\!\!,
\end{eqnarray}
where the variation of $e_J$ accounts for the corresponding uncertainty estimate.

Analogous to Refs.~\cite{Abbate:2010xh,Hoang:2014wka}, we have two additional scales whose functional form we briefly revisit. The first is the gap-subtraction scale $R(\rho)$:
\begin{eqnarray}
\label{eq:subtractionScale}
R(\rho) = \!\left\{\begin{array}{lrcl}
R(0) & 0 &\le& \rho < t_0
\\
\zeta(R(0),0,0,r_s \mu_H,t_0,t_1,\rho)\qquad &t_0 &\le& \rho < t_1
\\
\mu_S(\rho) & t_1 &\le& \rho \le 0.5
\end{array}
\right.\!\!.
\end{eqnarray}
The second is the non-singular renormalization scale $\mu_{\rm ns}$:
\begin{equation}\label{eq:NSScale}
\mu_{\rm ns}(\rho) = \mu_H - \frac{n_s}{2} \big[ \mu_H - \mu_J(\rho) \big]\,.
\end{equation}

Starting from $\rho=0.2$, we enter the region of the HJM spectrum where Sudakov shoulder logarithms are resummed. The shoulder resummation is also described by three scales: hard, jet and soft,
as summarized in Refs.~\cite{Bhattacharya:2022dtm,Bhattacharya:2023qet}. First of all, the shoulder hard scale is chosen to be the same as the dijet hard scale in Eq.~\eqref{eq:hardScale}. Since the resummation is performed in
position (Fourier) space, both jet and soft scales are set using the conjugate parameter $z$. Similar to the dijet region, we define a smooth transition function $g(\rho)$ to turn on the resummation in the shoulder region:
\begin{eqnarray}
\label{eq:gdef}
g(\rho) = \! \left\{\begin{array}{lrcl}
0 & 0 &\le& \rho < \rho_{L1}
\\
\zeta(0,0,1,0,\rho_{L1}, \rho_{L2},\rho)\qquad &\rho_{L1} &\le& \rho < \rho_{L2}
\\
1 \quad &\rho_{L2} &\le& \rho < \rho_{R1}
\\
\zeta(1,0,0,0, \rho_{R1},\rho_{R2},\rho) \qquad &\rho_{R1} &\le& \rho < \rho_{R2}
\\
0 & \rho_{R2} &\le& \rho < 0.5
\end{array}
\right.\!\!,
\end{eqnarray}
and the (hybrid) soft and jet scales read
\begin{equation}
\mu_S^{\text{sh}}(\rho,z)=\bigl[\mu_H\bigr]^{1-g(\rho)} \!\left[\sqrt{\left(2^{v_s}\frac{\mu_H e^{-\gamma_E}}{|z|}\right)^2+\left(\mu_S^{\text{min}}\right)^2}\;\right]^{g(\rho)},\qquad \mu_J^{\text{sh}}=\bigl[\mu_S^{\text{sh}}\bigr]^{v_j}\bigl[\mu_H\bigr]^{1-v_j}\,.
\end{equation}

Using Eq.~\eqref{eq:gdef}, the resummation scales are set to the canonical scales between $\rho_{L2}$ and $\rho_{R1}$. From $\rho_{L2}$ to $\rho_{L1}$ and from $\rho_{R1}$ to $\rho_{R2}$, we use the same quadratic function $\zeta(a_1,b_1,a_2,b_2,t_1,t_2,t)$ to gradually change them to the fixed-order scales. The values of shoulder profile parameters and their variations are also summarized in Table~\ref{tab:profParams}.
Note that we have also included the three-loop soft constant from the explicit calculation in Ref.~\cite{Baranowski:2024ysi}.
Compared to Ref.~\cite{Bhattacharya:2023qet}, we switch the sigmoid profile on the $\rho<1/3$ shoulder to the quadratic profile $\zeta(a_1,b_1,a_2,b_2,t_1,t_2,t)$, such that the shoulder resummation is completely turned off below $\rho_{L1}$. Consequently, we also change from the sigmoid profile parameters $\{\rho_L,\sigma_L\}=\{0.20,0.025\}$ to the quadratic profile parameters $\{\rho_{L1},\rho_{L2}\}=\{0.20,0.28\}$.

Lastly, we also implement a profile function for the non-perturbative power correction:
\begin{eqnarray}
\label{eq:npprof}
\Omega(\rho)= \! \left\{\begin{array}{lrcl}
\Om & 0 &\le& \rho < \rho_{L1}
\\
\zeta(\Om,0,\Th,0,\rho_{L1}, \rho_{L2},\rho)\qquad &\rho_{L1} &\le& \rho < \rho_{L2}
\\
\Th \quad &\rho_{L2} &\le& \rho < 0.5
\end{array}
\right.\!\!,
\end{eqnarray}
where we use the same smooth function $\zeta$ to connect the dijet $\Om$ and trijet $\Th$ power corrections as in Eq.~\eqref{eq:softScale}. We emphasize that we impose a positive $\Om$ but leave the sign of $\Th$ to be determined by the fit.

\vspace*{0.1cm}

\begin{figure}[t]
\centering
\begin{minipage}{0.48\textwidth}
\centering
\includegraphics[width=\textwidth]{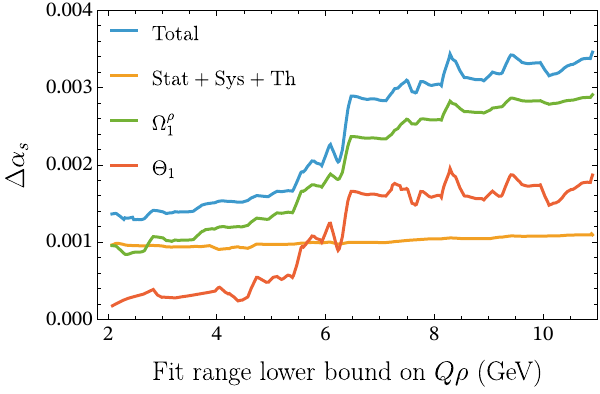}
\end{minipage}
\hfill
\begin{minipage}{0.48\textwidth}
\centering
\includegraphics[width=\textwidth]{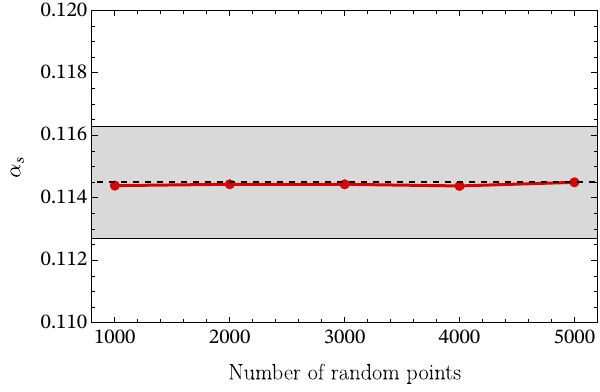}
\end{minipage}
\caption{Left panel: Contributions to the total uncertainty as a function of the lower bound on the fit range for the full theory prediction (dijet + shoulder + $\Om$ + $\Th$).
Right panel: In red we show the dependence of the best-fit value for $\alpha_s$ on the number of random profiles included in the scan in our best-fit setup, with the fit range $6\,{\rm GeV}/Q \le \rho \le 0.3$. The black dashed line and gray band correspond to the best-fit value and uncertainty with all 5000 profiles combined.}
\label{fig:errorBudgetRandomScan}
\end{figure}

{\bf Uncertainty decomposition:}
The overall uncertainty on our final value for the strong coupling
consists of three distinct parts: First, the combined experimental and perturbative theoretical uncertainties, where the former contains both statistical and systematic contributions. Second, the uncertainty induced by varying $\Om$, and third, the uncertainty induced by varying $\Th$. The contributions of these to the total uncertainties, as a function of the lower bound of the fit range, is displayed in the left panel of Fig.~\ref{fig:errorBudgetRandomScan}. When including more smaller-$\rho$ data, the fit uncertainty (yellow curve) is similar in size to the uncertainty induced by $\Om$ (green curve), whereas the uncertainty arising from $\Th$ (red curve) is comparatively small. This follows since, in most of the fit region, non-perturbative corrections mainly come from $\Om$. Conversely, if cutting at higher values of $a/Q$, one expects an increasing contribution of $\Th$ to the uncertainty budget, as the non-perturbative corrections arising from $\Th$ become more important. This behavior is evident in the left panel of Fig.~\ref{fig:errorBudgetRandomScan}. In addition, we observe an almost linear rise of the uncertainty tied to $\Om$ when increasing the lower bound of the fit range, which simultaneously yields a linear rise for the overall uncertainty on the strong coupling. Interestingly, the contribution from the fit uncertainty is rather flat, which suggests that the increased experimental uncertainty, due to less available data, is compensated for by a decrease of the perturbative theory uncertainty.

{\bf Random scans:} In the right panel of Fig.~\ref{fig:errorBudgetRandomScan} we show (in red) the fit result for the strong coupling as a function of the number of points used in the flat random scan. The fluctuation of the $\alpha_s$ central values stays well within our final fit uncertainty, displayed as the gray horizontal band.
In general, we find that at least 3000 points are needed to reliably saturate the $n$-dimensional parameter space. In all our analyses we took the conservative approach of using 5000 points for the random scan, suppressing fluctuations in this regard and obtaining robust results throughout. Many ways of incorporating correlations among theory parameters have been proposed (see~\cite{Cacciari:2011ze,Becher:2011fc,David:2013gaa,Bagnaschi:2014wea,Bonvini:2020xeo,Duhr:2021mfd,Tackmann:2024kci} and references therein) but the flat random scan offers the benefit of being both easy to implement and robust.

\begin{figure}[h!]
\centering
\begin{tikzpicture}[baseline]
\node[anchor=south west, inner sep=0] (image) at (0,0)
{\includegraphics[width=0.32\linewidth]{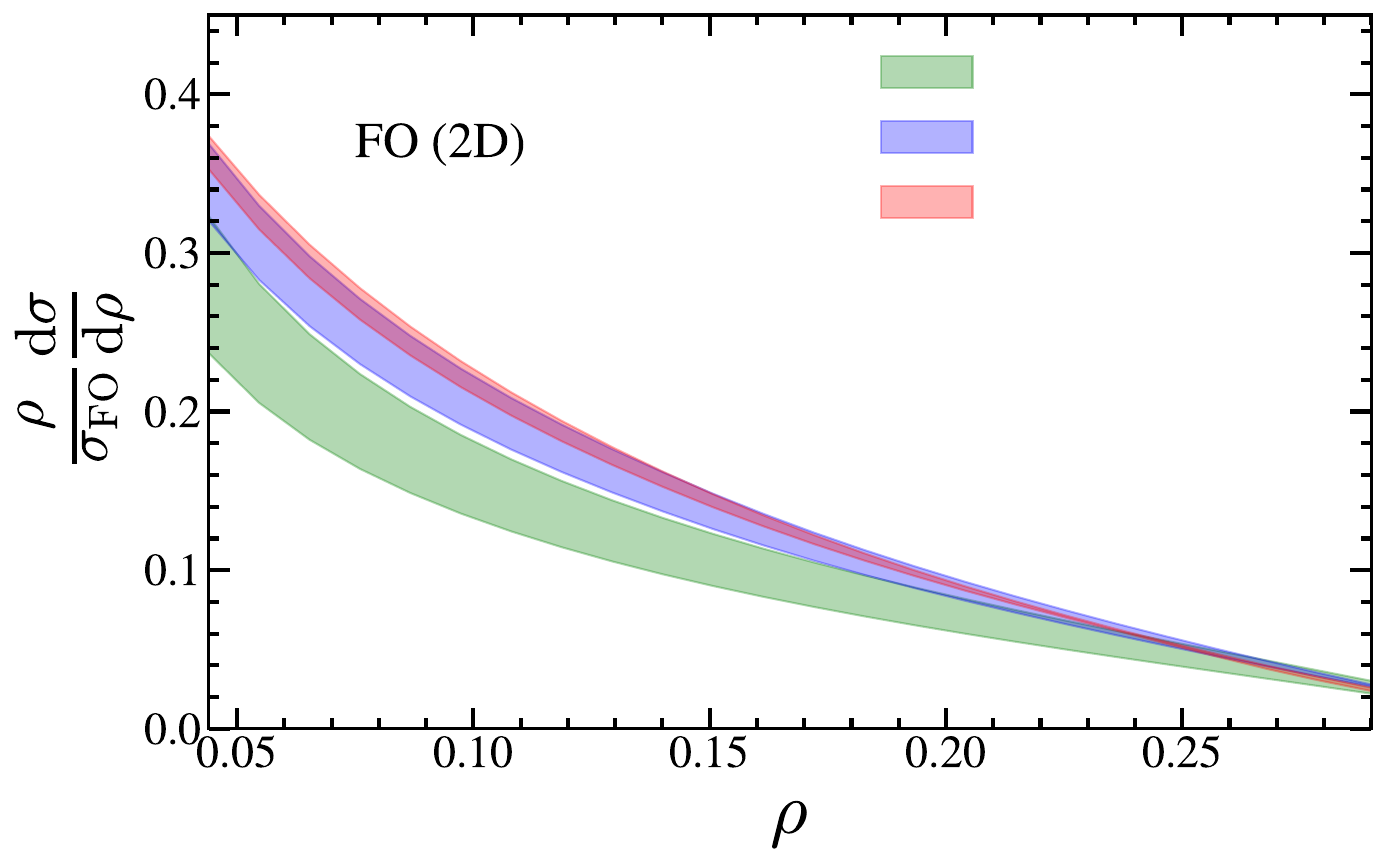}};
\begin{scope}[x={(image.south east)}, y={(image.north west)}]
\node[black, anchor=west,scale=0.6] at (0.71, 0.91) {LO};
\node[black, anchor=west,scale=0.6] at (0.71, 0.837) {NLO};
\node[black, anchor=west,scale=0.6] at (0.71, 0.763) {NNLO};
\end{scope}
\end{tikzpicture}
\begin{tikzpicture}[baseline]
\node[anchor=south west, inner sep=0] (image) at (0,0)
{\includegraphics[width=0.32\linewidth]{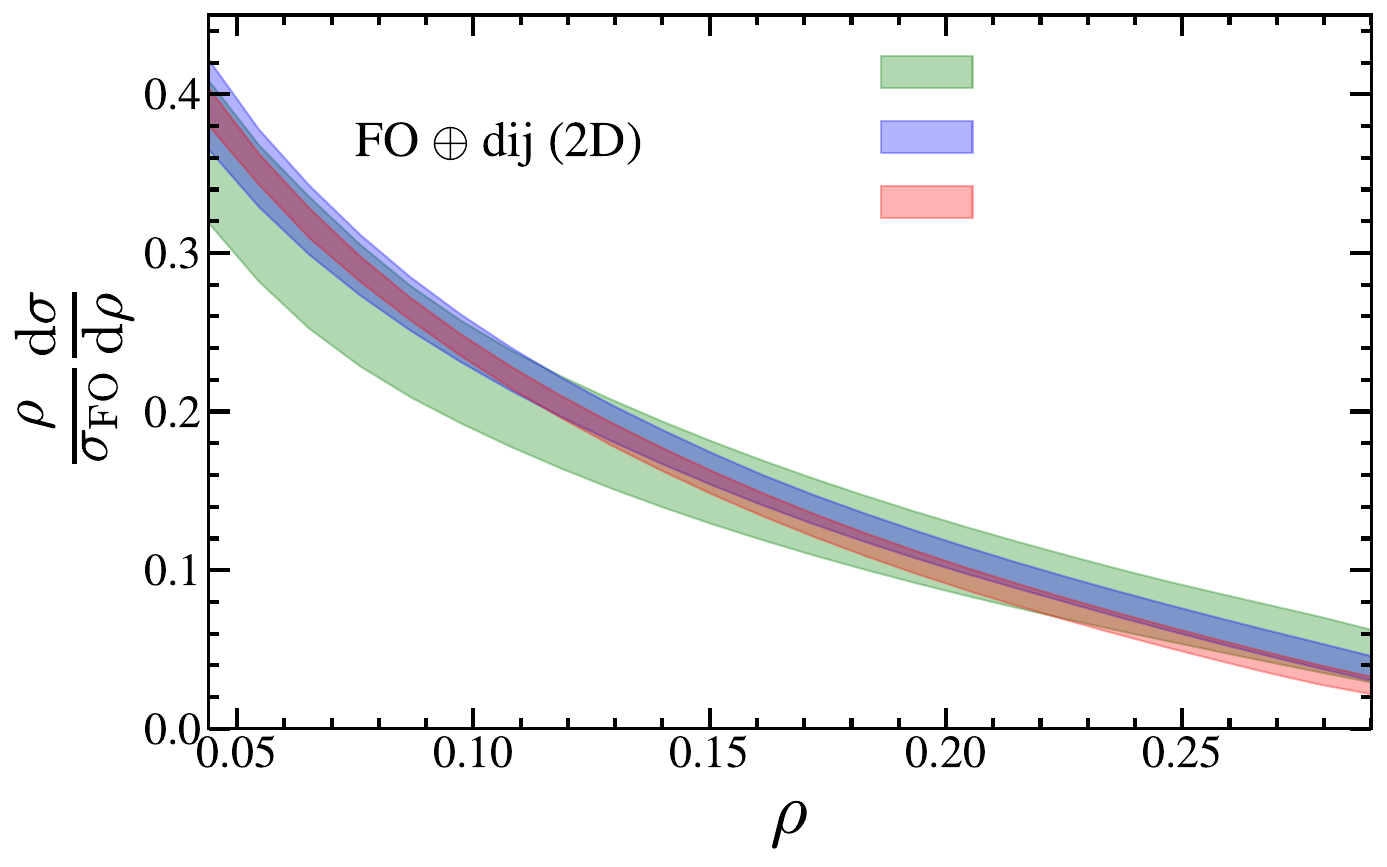}};
\begin{scope}[x={(image.south east)}, y={(image.north west)}]
\node[black, anchor=west,scale=0.6] at (0.71, 0.91) {NLL$^\prime$ +
LO};
\node[black, anchor=west,scale=0.6] at (0.71, 0.837) {N$^2$LL$^\prime$ + NLO};
\node[black, anchor=west,scale=0.6] at (0.71, 0.763) {N$^3$LL$^\prime$ + NNLO};
\end{scope}
\end{tikzpicture}
\begin{tikzpicture}[baseline]
\node[anchor=south west, inner sep=0] (image) at (0,0)
{\includegraphics[width=0.32\linewidth]{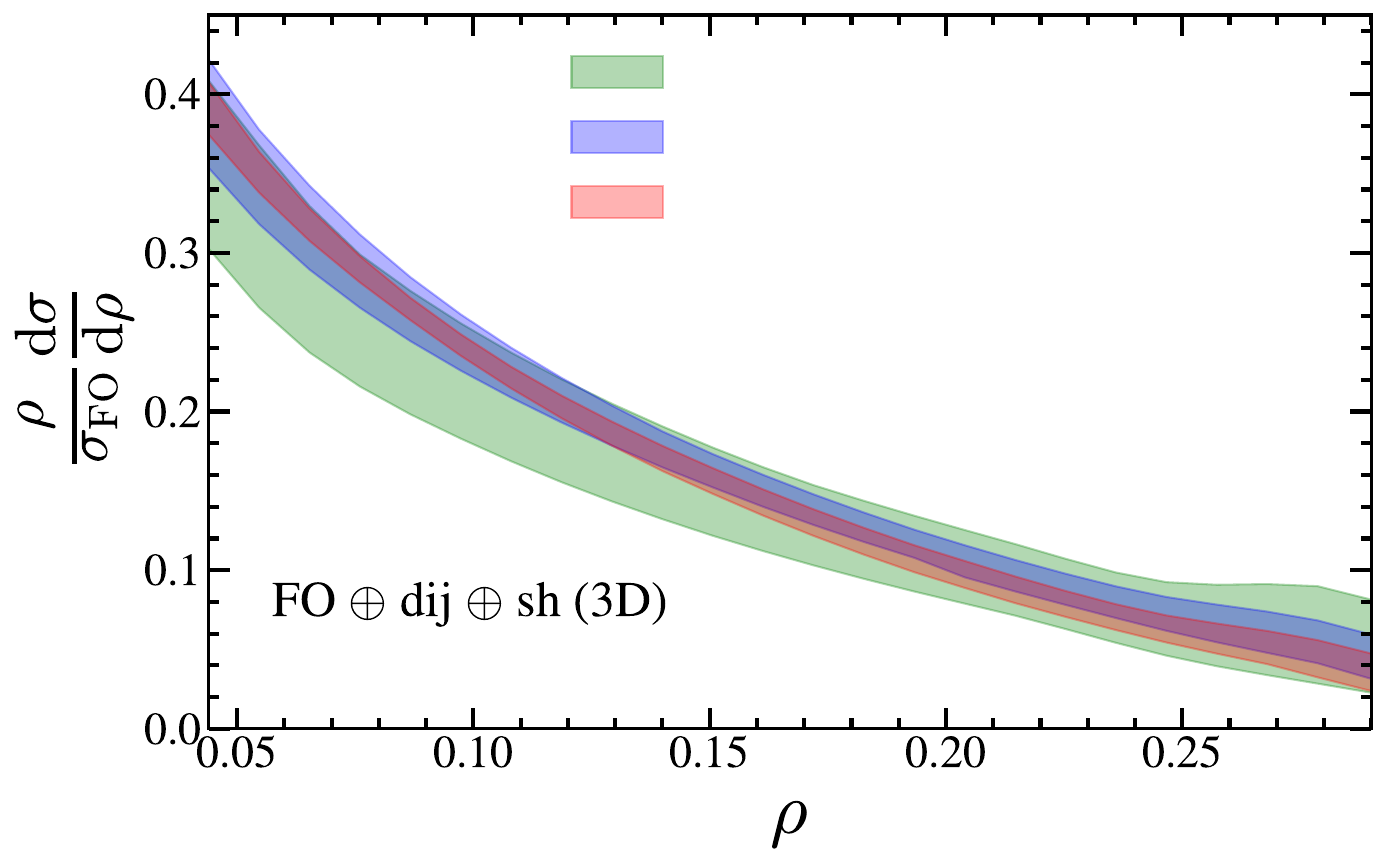}};
\begin{scope}[x={(image.south east)}, y={(image.north west)}]
\node[black, anchor=west,scale=0.6] at (0.49, 0.91) {NLL$^\prime$ + NLL$_{\rm sh}$ + LO};
\node[black, anchor=west,scale=0.6] at (0.49, 0.837) {N$^2$LL$^\prime$ + N$^2$LL$_{\rm sh}$ + NLO};
\node[black, anchor=west,scale=0.6] at (0.49, 0.763) {N$^3$LL$^\prime$ + N$^2$LL$_{\rm sh}$ + NNLO};
\end{scope}
\end{tikzpicture}
\caption{Convergence order-by-order in perturbation theory for FO (2D), FO+dij (2D) and FO+dij+sh (3D) and fixed values of the fit parameters. LO, NLO and NNLO denote the $\mathcal{O}(\alpha_s)$, $\mathcal{O}(\alpha_s^2)$ and $\mathcal{O}(\alpha_s^3)$ FO predictions, respectively.}
\label{fig:convergence}
\end{figure}

{\bf Convergence:}
In Fig.~\ref{fig:convergence} we show HJM theory predictions for various levels of sophistication at different orders in perturbation theory, including scale variations. The left panel displays the fixed-order distribution, and the central panel includes dijet resummation. The latter yields substantially better convergent behavior. Note, however, that in the central panel near $\rho\approx 0.25$ there is visibly less overlap between the different orders. This behavior is to a large extent remedied when including shoulder resummation, as shown in the right panel. This demonstrates the convergent character of our best theory description, which includes dijet and shoulder resummation.

Order-by-order convergence of the cross section is expected to translate into order-by-order convergent fit results.
Table~\ref{tab:convergenceFitResults} shows the difference between fits using the R-gap and $\overline{\rm MS}$ definitions for $\Omega_1$ at different orders in perturbation theory. Since this is a question associated with the dijet factorization we carry out fits for the simpler setup of the 2D fits using FO+dijet resummation. We see that the convergence in the two schemes is similar, while a slightly reduced fit uncertainty for $\alpha_s(m_Z)$ is achieved in the R-gap scheme. This can be contrasted with the case of thrust~\cite{Abbate:2010xh,Benitez:2024nav}, where the reduction in uncertainty from the R-gap scheme was more significant.

\begin{table}[t!]
\begin{tabular}{|l|c|c|c|c|c|c|}
\hline
& \multicolumn{3}{c|}{R-gap scheme} & \multicolumn{3}{c|}{$\msbar$ scheme}\\\hline
~Order~& $\alpha_s(m_Z)$ &
$\Om\,$[GeV] &
~$\chi^2/{\rm dof}$~ & $\alpha_s(m_Z)$ &
$\bar\Omega_1^\rho \,$[GeV] &
~$\chi^2/{\rm dof}~$\\
\hline
~N$^3$LL$^\prime$ & ~$0.1138\pm0.0015$~ & ~$0.65\pm 0.06$~ & $0.93$ & ~$0.1145 \pm 0.0019$~ & ~$0.60 \pm 0.06$~ &$0.92$\\
~N$^2$LL$^\prime$ & $0.1166 \pm 0.0027$ & $0.52 \pm 0.16$ &$ 0.94$ & $0.1178 \pm 0.0031$ & $0.49 \pm 0.16$ &$ 0.93$\\
~NLL$^\prime$ & $0.1157 \pm 0.0028$ & $0.85 \pm 0.12$ & $0.98$ & $0.1157 \pm 0.0030$ & $0.96 \pm 0.10$ & $0.99$\\ \hline
\end{tabular}
\caption{\label{tab:convergenceFitResults} Results of our 2D FO+dijet fits to the strong coupling and the leading power correction,
both in the R-gap and $\msbar$ schemes. The last three rows show the results for N$^3$LL$^\prime$ + $\ord{\alpha_s^3}$, N$^2$LL$^\prime$ + $\ord{\alpha_s^2}$, and NLL$^\prime$ + $\ord{\alpha_s}$ accuracy, respectively. Each determination is obtained from a fit to the dataset $\rho \in [4\,{\rm GeV}/Q,0.2]$. For each of the fit parameters at every order and scheme we display the central value and the fit uncertainty (the fit range and fit method uncertainties are not shown).}
\end{table}

\begin{figure}[h!]
\centering
\includegraphics[width=0.48\textwidth]{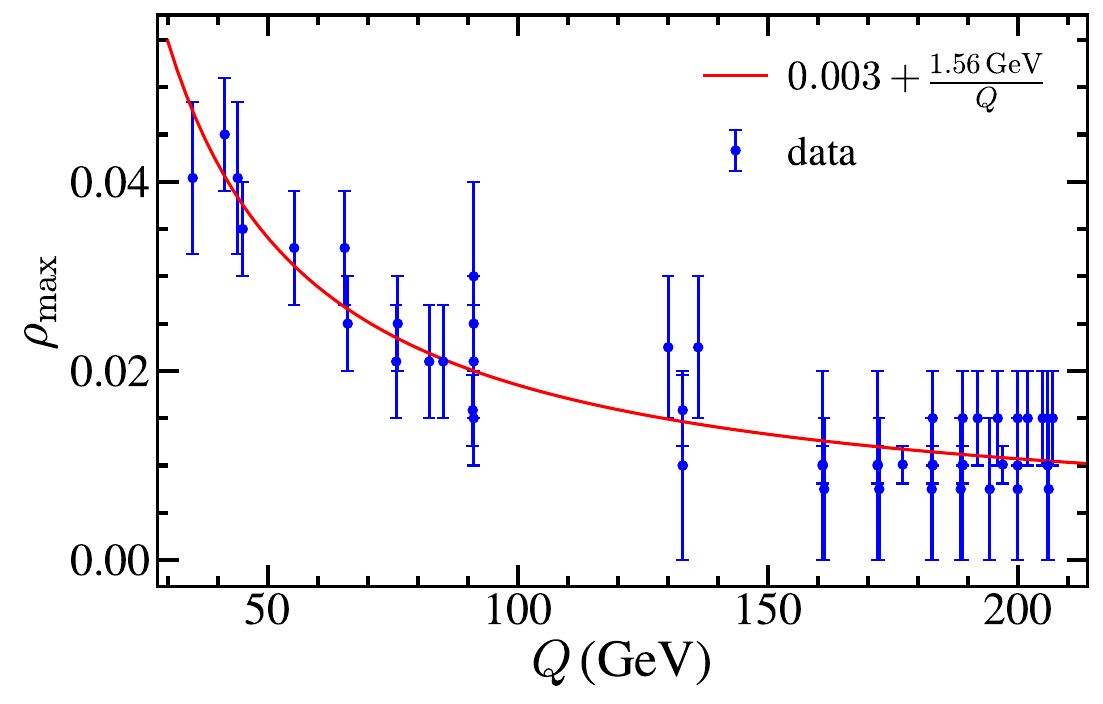}
\caption{Location of peaks among the 50 available datasets (blue points), where the error bars show the bin size of the highest point in the differential cross section. These are well reproduced by the red curve given by $\rho_\text{peak} = 0.003 + 1.56\,{\rm GeV}/Q$.}
\label{fig:peaks}
\end{figure}

\vspace*{2mm}
{\bf Peak region:}
For the lower bound of our fit range we use $a/Q$, in agreement with the argument from Ref.~\cite{Abbate:2010xh} for thrust that the peak position and increased sensitivity to non-perturbative effects goes as $1/Q$. That the peak position for HJM also goes as $1/Q$ is shown in Fig.~\ref{fig:peaks}, which displays the peak locations for for $50$ different datasets, represented by the blue dots.
The error bars indicate the bin size of the highest point in the differential distribution for the given center-of-mass energy. The red curve corresponds to the function $\rho_\text{peak} = 0.003 + 1.56\,{\rm GeV}/Q$. Our choice for the lower bound on the fit range, $a/Q$, is always taken to be well above these values.

\begin{figure}[t!]
\centering
\subfigure{\includegraphics[width=0.22\textwidth]{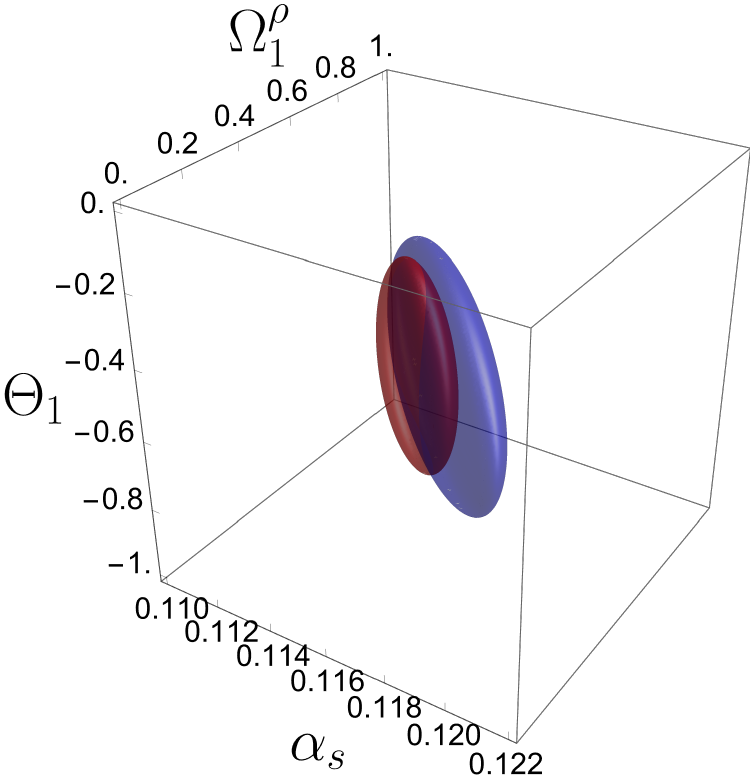}}
\subfigure{\includegraphics[width=0.225\textwidth]{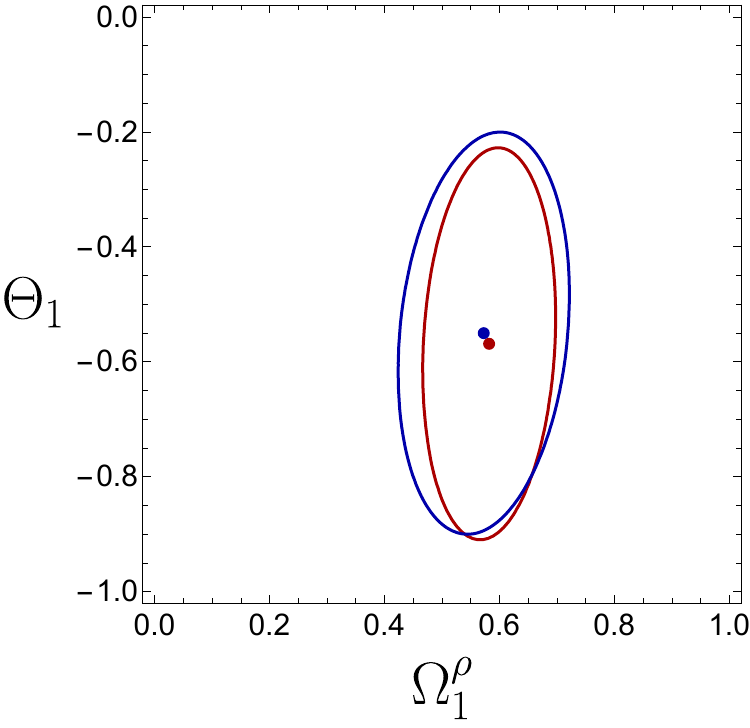}}
\subfigure{\includegraphics[width=0.23\textwidth]{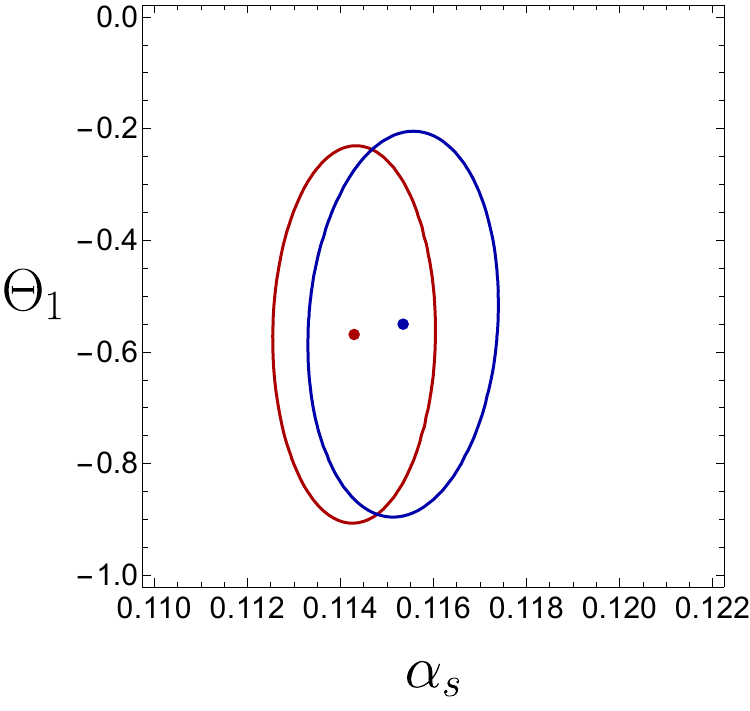}}
\subfigure{\includegraphics[width=0.22\textwidth]{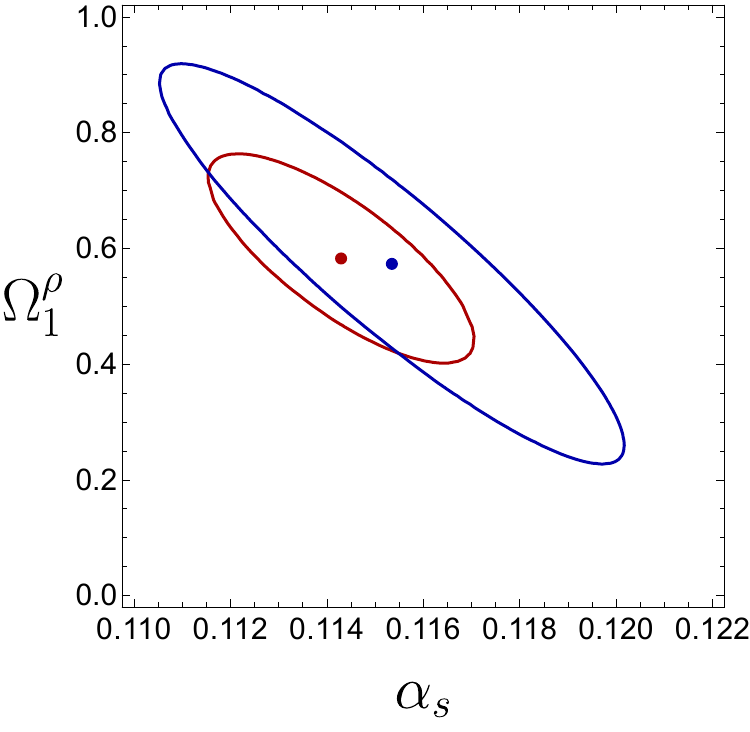}}
\caption{Correlations among the fit parameters $\alpha_s$, $\Om$ and $\Th$. Ellipses are 1-$\sigma$ confidence intervals ($\Delta \chi^2 = 3.53$) shown for the fit range $Q \rho > 3 \apeak$ (red) and $Q \rho > 6 \apeak$ (blue).}
\label{fig:ellipses}
\end{figure}
\vspace*{0.25cm}

\begin{table}[b!]
\centering
\begin{tabular}{|c|c|c|c|c|c|c||c|c|}
\hline
Model & $\alpha_s(m_Z)$ & th+exp
&
$\Om$ &
$\Th$ & fit range
& ~$\chi^2$/dof~ & $\Om$\,[GeV] & $\Th$\,[GeV] \\
\hline
Fixed Order 2D & ~$0.1166^{+0.0037}_{-0.0033}$~ & ~\makecell{$+0.0013$\\$-0.0013$}~ & ~\makecell{$+0.0031$\\$-0.0026$}~ & -- & ~$\pm\,0.0015$~ & $1.108$ & $0.06 \pm 0.11$ & -- \\
FO\,+\,dijet 2D & $0.1149_{-0.0015}^{+0.0010}$ & ~\makecell{$+0.0005$\\$-0.0008$} & ~\makecell{$+0.0009$\\$-0.0013$} & -- &$\pm\,0.0001$ & $1.044$ & $0.56 \pm 0.06$ & -- \\
FO\,+\,dijet 3D & $0.1152^{+0.0026}_{-0.0021}$ & ~\makecell{$+0.0009$\\$-0.0009$} & ~\makecell{$+0.0022$\\$-0.0017$} & ~\makecell{$+0.0005$\\$-0.0003$}~ & $\pm\,0.0009$ & $1.055$ & $\,0.56 \pm 0.09\,$ & $~\,\,0.59 \pm 0.13$ \\
\,FO\,+\,dijet\,+\,shoulder 3D\, & $0.1148^{+0.0013}_{-0.0020}$ & ~\makecell{$+0.0005$\\$-0.0010$} & ~\makecell{$+0.0011$\\$-0.0017$} & ~\makecell{$+0.0002$\\$-0.0001$} &$\pm\,0.0005$ & $1.039$ & $0.61 \pm 0.08$ & \,$-0.46 \pm 0.17$\, \\
\hline
\end{tabular}
\caption{Fits for $\alpha_s,\Om$ and $\Th$ for different levels of theoretical sophistication. These numbers come from
a weighted average of different fit-range lower bounds as described in the main text. The lower bound is taken as $c \,\apeak/Q$ with $c=5-8$ in steps of $0.5$ for FO 2D, and $c=3-6$ in steps of $0.5$ for the other three rows. (If FO 2D is taken with the same range as the other approximations, then its fit range uncertainty is even larger.)
\label{tab:fits}}
\end{table}

{\bf Correlation among fit parameters:}
Fig.~\ref{fig:ellipses} displays the correlations among fit parameters for our best fit setup. The left panel shows the correlation among $\alpha_s$, $\Om$ and $\Th$ as an ellipsoid, whereas the remaining panels display its 2-dimensional orthographic projections. In particular, the second panel gives the correlation between $\Th$ and $\Om$, the third panel between $\Th$ and $\alpha_s$, and the final panel between $\Om$ and $\alpha_s$.
The strongest correlation is between $\Om$ and $\alpha_s$, which is also reflected in the fit result shown in Eq.~(\ref{fit}), where the largest contribution to the overall uncertainty of $\alpha_s$ comes from the uncertainty induced by $\Om$.

\vspace*{2mm}
{\bf Fit components:} The breakdown in uncertainties for the fits in Fig.~\ref{fig:alphasfits} is given in Table~\ref{tab:fits}.
For fixed order, the lower bound of the fit range is varied from ${5 \apeak}/{Q}$ to ${8 \apeak}/{Q}$ while for the other fits it is varied from ${3 \apeak}/{Q}$ to ${6 \apeak}/{Q}$. Increasing the lower bound reduces the fit-range error for FO, but leads to a larger error from the NP parameter $\Om$.
With resummation, there is little sensitivity to the lower bound so it can be taken low, reducing the uncertainty from $\Om$. If the lower bound is taken too close to $\apeak/Q$,
higher order dijet power corrections of ${\cal O}(\Lambda^2_{\rm QCD})$ and beyond become leading order, necessitating a full non-perturbative shape function, so we always keep it above $3 \apeak/Q$.

\vspace*{2mm}
{\bf $\chi^2$ results:} Supplemental to Fig.~\ref{fig:lowercut}, we also provide the reduced $\chi^2$ for different fit range lower cuts in Fig.~\ref{fig:chi2dof}. Starting from pure fixed-order ($\text{FO (2D)}$), the $\chi^2/\text{dof}$ decreases as we include more theoretical components, with our most precise theory prediction ($\text{FO}\oplus\text{dij}\oplus\text{sh (3D)}$) having the lowest value, indicating that it provides the best description of the experimental data.

\begin{figure}[t!]
\centering
\includegraphics[width=0.5\textwidth]{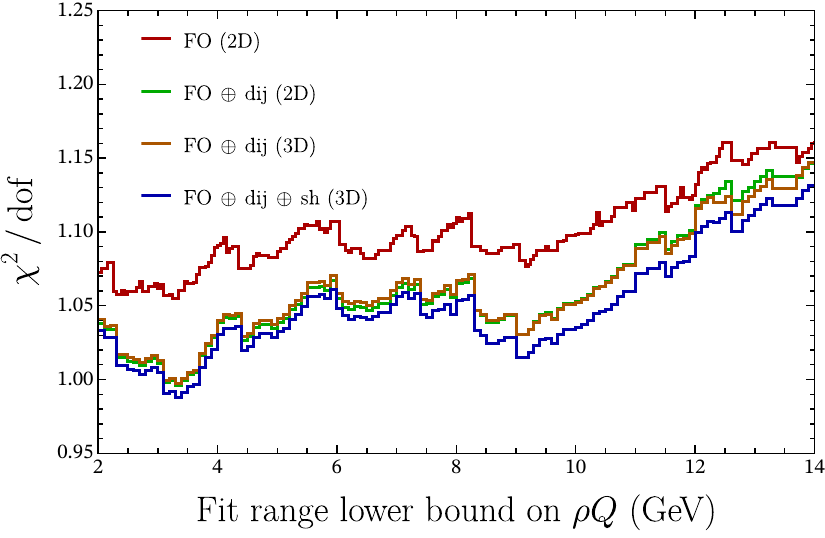}
\caption{$\chi^2/\text{dof}$ with respect to different lower cuts in the fit range. Our full theoretical setup, $\text{FO}\oplus\text{dij}\oplus\text{sh (3D)}$, leads to the lowest $\chi^2/\text{dof}$.}
\label{fig:chi2dof}
\end{figure}

\end{widetext}

\end{document}